\documentstyle[12pt,a4,epsfig]{article}
\renewcommand{\thefootnote}{\fnsymbol{footnote}}

\newcommand{\titul}[1] {\boldmath \begin{center}{\Large {\bf #1 } } 
\end{center}
\vskip 0.8cm}
\newcommand{\autor}[1] {\begin{center}  {\bf \lineskip .3cm #1  }
                        \end{center} }
\newcommand{\place}[1] {\begin{center}  {\normalsize \bf \it #1   } \end{center}}

\date{\today}
\title{GMM}
\date{\today}
\renewcommand{\thefootnote}{\fnsymbol{footnote}}

\date{\today}

\def\bmaT{\left(\begin{array}{ccc}}
\def\emaT{\end{array}\right)}
\def\bma{\left( \begin{array} }
\def\ema{\end{array} \right)}

\def\gsim{~{\rlap{\lower 3.5pt\hbox{$\mathchar\sim$}}\raise 1pt\hbox{$>$}}\,}
\def\lsim{~{\rlap{\lower 3.5pt\hbox{$\mathchar\sim$}}\raise 1pt\hbox{$<$}}\,}

\def\bea{\begin{eqnarray}}
\def\eea{\end{eqnarray}}
\def\nn{\nonumber}

\def\Journal#1#2#3#4{{#1} {\bf #2} (#4) #3}

\def\PLB{{ Phys. Lett.} B}

\begin{document}
\hbadness=10000
\pagenumbering{arabic}
\begin{titlepage}
%
\begin{center}
\titul{\bf  Unitarity bounds in the Higgs model including \\[2mm]triplet fields
 with custodial symmetry}
\autor{ 
Mayumi Aoki$^1$\footnote{mayumi@icrr.u-tokyo.ac.jp} and Shinya Kanemura$^2$\footnote{kanemu@sci.u-toyama.ac.jp}}
\place{1: ICRR, University of Tokyo, Kashiwa 277-8582, Japan}
\place{2: Department of Physics, University of Toyama, \\
3190 Gofuku, Toyama 930-8555, Japan}

\end{center}

\vskip2.0cm

\begin{abstract}
\noindent
 We study bounds on Higgs boson masses from perturbative unitarity
 in the Georgi-Machacek model, whose Higgs sector is composed of a scalar
 isospin doublet, a real and a complex isospin triplet fields.
 This model can be compatible with the electroweak precision data without fine tuning
 because of the imposed global $SU(2)_R$ symmetry in the Higgs
 potential, by which the electroweak rho parameter is unity at the tree level. 
 All possible two-body elastic-scattering channels are taken into
 account to evaluate the S-wave amplitude matrix, and 
 then the condition of perturbative unitarity is imposed on  
 the eigenvalues to obtain constraint on the Higgs parameters.
 Masses of all scalar bosons turn out to be bounded from above, some of which 
 receive more strict upper bounds as compared to that in the standard
 model (712 GeV).
 In particular, the upper bound of the lightest scalar boson,
 whatever it would be, is about 270 GeV.
 \end{abstract}

\vskip1.0cm
{\bf  PACS index :
12.60.Fr, 14.80.Cp
}
\end{titlepage}
\thispagestyle{empty}
\newpage
\pagestyle{plain}
\renewcommand{\thefootnote}{\arabic{footnote} }
\setcounter{footnote}{0}
\section{Introduction}

The nature of electroweak symmetry breaking remains unknown at the present
status of our knowledge for high energy physics.
In the standard model (SM), a scalar isospin doublet field, the Higgs 
field, is introduced to be responsible for spontaneous breakdown
of electroweak gauge symmetry.
Its vacuum expectation value~(VEV) triggers the symmetry
breaking, so that it provides origins of
masses of weak bosons via the Higgs mechanism,
and also does those of quarks and charged leptons via Yukawa
interaction.
Although the SM Higgs sector is simple, the Higgs sector could have
a more complicated structure in the actual world. In particular, when 
the Higgs sector  would play an additional role to explain phenomena
which the SM cannot, it should necessarily be
an extended form from the SM one. 
Therefore, experimental detection of the Higgs boson and precision
measurements of its properties are extremely important not only to
confirm our basic idea of electroweak symmetry breaking but also to
determine details of the Higgs sector and further to outline 
the structure of new physics.

In constructing an extended Higgs sector, there are two important
requirements from current experimental data.
First of all, 
the data indicate that the electroweak rho parameter ($\rho$) is very close to unity.
Second, flavor of quarks and charged leptons is (approximately)
conserved in the neutral current.
In the SM, these two conditions are satisfied respectively
by the custodial symmetry which ensures $\rho=1$
at the tree level, and by the Glashow-Iliopoulos-Maiani~(GIM) mechanism
which prohibits the tree-level flavor changing neutral current~(FCNC). 
Needless to say that these experimental requirements must be
respected in extended Higgs models which would appear in the low
energy effective theory of a more fundamental theory beyond the SM.

Extension of the SM Higgs sector can be considered by including
additional scalar isospin singlets, doublets and higher multiplets.
It is known that additional singlets and doublets
keep $\rho=1$ at the tree level~\cite{Gildener:1976ih}. Radiative corrections can slightly deviate the
rho parameter from unity, corresponding to explicit violation of the
custodial symmetry in the dynamics in the loop.
On the other hand, extension with higher multiplets such as triplets
is usually problematic, predicting the rho parameter
to be explicitly different from unity already at the tree level~\cite{Gunion:1992hs}.
One way to avoid this problem is to make a fine-tuning on the size of 
vacuum expectation values of the triplet fields; i.e., to set tiny values on them.
Another possibility is to impose the custodial symmetry to the Higgs sector,
so that the rho parameter is predicted
to be unity at the tree level. 
In 1985 Georgi and Machacek proposed such a model with one real triplet 
($Y$=0) and one complex triplet ($Y$=2) in addition to 
the Higgs doublet~\cite{Georgi:1985nv}. 
Chanowitz and Golden have explicitly constructed 
the Higgs potential of
this model~\cite{Chanowitz:1985ug}; i.e., 
imposing the custodial $SU(2)_{\rm V}$ symmetry to the potential, VEVs of all the
isospin triplets become common, and then the tree-level value of the rho parameter is unity. 
They also have shown that the quantum correction from the scalar sector is stabilized 
by such a global symmetry, so that the rho parameter is corrected at the loop 
level only due to explicit $SU(2)_{\rm V}$ violation in the other sectors such as
hypercharge interaction and Yukawa interaction, just like in the SM.
Several phenomenological studies have been done on this model in 
Refs.~\cite{Gunion:1989ci,Vega:1989tt,Gunion:1990dt,Godbole:1994np,Cheung:1994rp,
Akeroyd:1998zr,Haber:1999zh,Cheung:2002gd}.  

Generally in extended Higgs models, there are many free parameters in the Higgs potential, which
spoil predictive power of the model. Hence, it is important to clarify allowed
regions in the parameter space not only by using current experimental
data but also by investigating theoretical consistencies such as
perturbative unitarity~\cite{Lee:1977yc,Dicus:1992vj}, 
vacuum stability and triviality~\cite{Lindner:1985uk}.
This kind of study has been often developed to constrain parameters of the Higgs sector
in the context of the two-Higgs-doublet model~\cite{Kanemura:1993hm,
Ginzburg:2005dt, Komatsu:1981xh, Nie:1998yn}, and  in a specific triplet model~\cite{Forshaw:2003kh}. 

In this paper, we study bounds on Higgs boson masses  from perturbative unitarity  
in the Georgi-Machacek (GM) model. 
The Higgs potential respects the global $SU(2)_R$ symmetry,
so that the custodial $SU(2)_{\rm V}$ symmetry remains after the
electroweak symmetry breaking ($SU(2)_{L} \otimes SU(2)_R \to SU(2)_{\rm V}$). 
There are ten physical scalar states, which can be expressed by 
a $SU(2)_{\rm V}$ 5-plet $(H_5^{++}, H_5^+, H_5^0, H_5^-, H_5^{--})$,
a 3-plet $(H_3^+, H_3^0, H_3^-)$
and two singlets $\tilde{H}_1^0$ and $\tilde{H}^{0'}_{1}$~\cite{Georgi:1985nv}.
The scalar components in the same multiplet are
degenerate in mass at the tree level. 
In the Higgs potential of the GM model,
explicit $Z_2$ violation can only appear in the 
trilinear scalar interaction,
but they must be forbidden to avoid excessive magnitudes for masses of neutrinos.
Neglecting such terms by imposing the $Z_2$ symmetry, all Higgs boson masses in this
model are described in terms of the VEV, the mixing angles and 
the dimension-less coupling constants $\lambda_i$ in the Higgs potential.
This situation is somewhat similar to the two-Higgs-doublet model
with the discrete $Z_2$ symmetry~\cite{Glashow:1976nt},   
in which perturbative unitarity gives upper bounds on all the
Higgs boson masses~\cite{Kanemura:1993hm}.

In our analysis, all possible two-body elastic scattering channels
(91-channels) are taken into account
to evaluate the S-wave amplitude matrix in the GM model.
Constraints on the Higgs parameters are obtained by imposing
the condition of partial wave unitarity on 
the eigenmatrix of the S-wave amplitude. 
Masses of all scalar bosons turn out to be bounded from above, some of which
receive much stronger bounds as compared to that in the SM (712 GeV). 
In particular, the mass of 
at least one of the charged Higgs bosons should be less than about 400 GeV. 
At least one of the neutral Higgs boson is lighter than 322 GeV.
Furthermore, the upper bound of the lightest scalar boson,
whatever it would be, can be about 269 GeV.
We also find that by using the experimental constraints
from $Zb\bar b$ results~\cite{Haber:1999zh},
the combined upper bound for the lightest Higgs boson is
lower than 269 GeV, depending on what the lighest is. 
Therefore, the model can be well testable at current and future
collider experiments. 
 
In Sec.~2, a brief review of the GM model is given.
The transition matrix for two-body elastic scatterings is
calculated in the  high-energy limit, and its eigenmatrix is obtained in
Sec.~3.
In Sec.~4, the condition of S-wave unitarity is imposed for the
eigenmatrix of the transition matrix, and bounds on the Higgs boson
masses are evaluated.
Conclusions are presented in Sec.~5.

\section{Georgi-Machacek Model}

The GM model contains a complex $SU(2)_L$ doublet field $\phi$ ($Y$=1), a real
$SU(2)_L$ triplet field $\xi$ ($Y$=0) and a complex $SU(2)_L$ triplet
field $\chi$ ($Y$=2)~\cite{Georgi:1985nv}, and respects the global $SU(2)_R$ symmetry in the Higgs potential~\cite{Chanowitz:1985ug} . They can be described by the form of $SU(2)_L\otimes SU(2)_R$ 
multiplets $\Phi$ and $\Delta$ in the potential; 
\bea
\Phi=\left(\begin{array}{cc} \phi^{0\ast}&\phi^{+} \\ 
\phi^- & \phi^0 \end{array}\right),~~~\Delta=\left(\begin{array}{ccc} \chi^{0\ast}&\xi^+&\chi^{++} \\ 
\chi^- & \xi^0 &\chi^+ \\\chi^{--} &\xi^- &\chi^0\end{array}\right) \ ,
\eea
where $\phi=(\phi^+, \phi^0)^T$, $\xi=(\xi^+, \xi^0, \xi^-)^T$ and 
$\chi=(\chi^{++}, \chi^+, \chi^0)^T$, and
$\phi^-=-(\phi^+)^\ast$, $\xi^-=-(\xi^+)^\ast$ and $\chi^-=-(\chi^+)^\ast$~\cite{Gunion:1989ci}.
The most general Higgs potential is given by
\bea
V&=&m_1^2{\rm Tr}(\Phi^\dag\Phi)+m_2^2{\rm Tr}(\Delta^\dag\Delta)+\lambda_1{\rm Tr}(\Phi^\dag\Phi)^2+\lambda_2{\rm Tr}(\Delta^\dag\Delta)^2+\lambda_3{\rm Tr}(\Phi^\dag\Phi){\rm Tr}(\Delta^\dag\Delta) \nn \\
&&+\lambda_4{\rm Tr}(\Delta^\dag\Delta\Delta^\dag\Delta)+\lambda_5{\rm Tr}(\Phi^\dag\frac{\tau_i}{2}\Phi\frac{\tau_j}{2}){\rm Tr}(\Delta^\dag T_i\Delta T_j) \nn \\
&&+\mu_1{\rm Tr}(\Phi^\dag\frac{\tau_i}{2}\Phi\frac{\tau_j}{2})\Delta_P^{ij}+\mu_2{\rm Tr}(\Delta^\dag T_i \Delta T_j)\Delta_P^{ij}\ ,
\label{potential}
\eea
where $\tau_i$ are the $2\times 2$ Pauli matrices and 
\bea
\Delta_P=P^\dag\Delta P,~~~~~~~~~P=\left(\begin{array}{ccc}-1/\sqrt{2}&i/\sqrt{2}&0\\0&0&1\\1/\sqrt{2}&i/\sqrt{2}&0\end{array}\right)\ .
\eea

The neutral components of the doublet and the real and the complex triplets have the VEVs, $v_\phi$, $v_\xi$, and $v_\chi$, respectively, which are defined as 
\bea
\phi^0&=&\frac{v_\phi+\phi_r^0+i\phi_i^0}{\sqrt{2}}\ , \\
\xi^0&=&v_\xi+\xi_r^0\ , \\
\chi^0&=&v_\chi+\frac{\chi_r^0+i\chi_i^0}{\sqrt{2}} \ .
\eea
After electroweak symmetry breaking, the custodial $SU(2)_{\rm V}$ symmetry remains in the Higgs sector $(SU(2)_L\times SU(2)_R\to SU(2)_{\rm V})$, by which 
the real and complex triplets have the same VEV, $v_\Delta\equiv v_\xi=v_\chi$.
Consequently this leads to $\rho=1$ at the tree level~\cite{Chanowitz:1985ug}.
In this case the VEVs are constrained as $v^2=v_\phi^2 + 8v_\Delta^2$,
where $v=(\sqrt{2}
G_F)^{-\frac{1}{2}} \simeq 246$ GeV.
Therefore, differently from usual triplet models, $v_\Delta$ can be
 of order 100 GeV in this model without explicit inconsistency with the
 experimental value of the rho parameter. 
It is convenient to introduce the doublet-triplet mixing angle $\theta_H$,
\bea
\tan \theta_H= \frac{2\sqrt{2}v_\Delta}{v_\phi}\ .
\eea
The experimental constraint on $\theta_H$ is discussed in Ref.~\cite{Haber:1999zh}.

In the potential Eq.(\ref{potential}),
the last two terms with the coupling constants $\mu_1$ and $\mu_2$ explicitly 
violate the discrete $Z_2$ symmetry under the transformation of
$\Phi\to \Phi$ and $\Delta \to -\Delta$. 
Without the $Z_2$ symmetry, the model is allowed to have the mass terms for the neutrinos 
by assigning of lepton number $-2$ to the complex triplet field,
\bea
i(h_\nu)_{ab}\psi_{La}^TC\tau_2\hat\chi\psi_{Lb}+{\rm h.c.},
\eea 
where $\hat \chi=\frac{\tau^i}{2}(P^\dag\chi)^i$.
In order to generate the tiny neutrino masses the Yukawa coupling $h_\nu$ should be fine-tuned to be very small as $h_\nu\sim {\cal O}(10^{-12})$
for the triplet VEV of order 100 GeV.
Since we would like to avoid such fine tuning with respect to the neutrino masses,
we require the discrete $Z_2$ symmetry in the Higgs potential and prohibit the last two terms in Eq.(\ref{potential})\footnote{
The neutrino masses might be generated by any other mechanism
(e.g. \cite{Chun:2003ej}).
We shall discuss it elsewhere~\cite{Aoki-Kanemura:neutrino}. 
}.
Therefore, quarks and leptons couple to the $SU(2)_L$ doublet field
$\Phi$ in the same way as the SM Yukawa coupling, but do not to
the triplet $\Delta$ at the tree level. Because all the masses
of quarks and leptons are obtained from the VEV in $\Phi$, we do not
have to worry about FCNC, and it is expected to appear at most at the same level as
that in the SM.
This property of the coupling to fermions would give an additional
constraint on the value of the doublet-triplet mixing angle
$\theta_H^{}$ by $\tan\theta_H^{}< {\cal O}(1)$, since large
values of $\tan\theta_H^{}$ $(\gg 1)$ imply that the top-Yukawa coupling
is much greater than ${\cal O}(1)$.

In the GM model, there are ten physical states in the Higgs sector, 
which are classified as a 5-plet ($H_5^{++}, H_5^{+}, H_5^{0}, H_5^{-}, H_5^{--}$), a 3-plet
($H_3^{+}, H_3^{0}, H_3^{-}$), and two singlets $H_1^0$ and $H_1^{0'}$ under the custodial 
$SU(2)_{\rm V}$ symmetry.
These are given in terms of the original component fields and the doublet-triplet mixing angle $\theta_H$ as~\cite{Gunion:1989ci}
\bea
&&H_5^{++}=\chi^{++}\ , \\
&&H_5^{+}=(\chi^{+}-\xi^{+})/\sqrt{2} \ ,\\
&&H_5^{0}=(2\xi_r^0-\sqrt{2}\chi^{0}_r)/\sqrt{6}\ , \\
&&H_3^{+}=\cos{\theta_H}(\chi^++\xi^+)/\sqrt{2}-\sin{\theta_H}\phi^+\,, \\
&&H_3^{0}=i(-\cos{\theta_H}\chi^0_i+\sin{\theta_H}\phi^0_i) \,, \\
&&H_1^{0}=\phi_r^0\,, \\
&&H_1^{0'}=(\sqrt{2}\chi^0_r+\xi_r^0)/\sqrt{3}\,. 
\eea
The 5-plet components do not include the component fields from
the isospin doublet field $\Phi$, 
so that the states of the 5-plet do not couple to the fermions at the tree level. 
On the other hand, the 3-plet fields can couple to the fermions.
Because of invariance under the custodial $SU(2)_{\rm V}$ symmetry, states in the different multiplet cannot mix each other.

All members in the same $SU(2)_{\rm V}$ multiplet are degenerate in mass at the tree level. 
The masses of the 5-plet and the 3-plet are respectively given by  
\bea
m_{H_5}^2&=&(\lambda_4\sin^2\theta_H-\frac{3}{2}\lambda_5\cos^2\theta_H) v^2 \ ,
\label{m5} \\
m_{H_3}^2&=&-\frac{\lambda_5}{2}v^2 \ .
\eea
On the other hand, two $SU(2)_{\rm V}$ singlets can mix, and the mass matrix 
\bea
{\cal M}^2_{H_1^0,H_1^{0'}}=\left(\begin{array}{cc}8\cos^2\theta_H\lambda_1
&\sqrt{\frac{3}{2}}\sin\theta_H\cos\theta_H  (2\lambda_3+\lambda_5)   \\
\sqrt{\frac{3}{2}}\sin\theta_H\cos\theta_H  (2\lambda_3+\lambda_5)   
&\sin^2\theta_H(3\lambda_2+\lambda_4)\end{array}\right)v^2\, 
\label{masses}
\eea
is diagonalized by introducing the mixing angle $\alpha$.
The eigenvalues correspond to 
the masses $m_{\tilde H_1^0}$ and $m_{\tilde H_1^{0'}}$ for 
the mass eigenstates ${\tilde H_1^0}$ and ${\tilde H_1^{0'}}$.

From Eqs.(\ref{m5}) - (\ref{masses}), the quartic couplings $\lambda_i$ are expressed in terms of the masses and the mixing angles as
\bea
\lambda_1&=&(m_{\tilde H_1^0}^2\cos^2\alpha+m_{\tilde H_1^{0'}}^2\sin^2\alpha)/(8v^2\cos^2\theta_H)\ , 
\label{lambda1}\\
\lambda_2&=&(m_{\tilde H_1^0}^2\sin^2\alpha+m_{\tilde H_1^{0'}}^2\cos^2\alpha-m_{H_5}^2+3m_{H_3}^2\cos^2\theta_H)/(3v^2\sin^2\theta_H)\ , 
\label{lambda2}\\
\lambda_3&=&(m_{\tilde H_1^{0'}}^2-m_{\tilde H_1^0}^2)\cos\alpha\sin\alpha/(\sqrt{6}v^2\sin\theta_H\cos\theta_H)+m_{H_3}^2/v^2\ , 
\label{lambda3}\\
\lambda_4&=&(m_{H_5}^2-3m_{H_3}^2\cos^2\theta_H)/(v^2\sin^2\theta_H)\ , \\
\lambda_5&=&-2m_{H_3}^2/v^2 \ .
\label{lambda5}
\label{lambda}
\eea

The $SU(2)_{\rm V}$ 3-plet fields receive the constraints from the current data of $Z\to b\bar b$, $B_0-\bar B_0$ and $K_0-\bar K_0$ mixings~\cite{Kundu:1995qb, Chakraverty:1995zw}.
These data give bounds on the mass $m_{H_3}^{}$ with the mixing angle $\theta_H$.
The most stringent experimental constraint comes from $Z \to b \bar b$.
The mass $m_{H_3}^{}$ is constrained to be smaller than 1 (0.5) TeV for $\tan\theta_H~<$ 2 (1).

Although the 5-plet fields do not couple to the fermions, the singly-charged state in the 5-plet has a characteristic coupling of $H_5^\pm W^\mp Z$, which only appears beyond the tree level
in multi-Higgs-doublet models~\cite{Grifols:1980uq}.
Experimental confirmation of a sizable coupling of $H_5^\pm W^\mp Z$ 
with $\rho \simeq 1$ should be a strong indication for the GM
model\cite{Mukho:1990}. 
This coupling is testable via the process  $p\bar p\to W^\pm H^\mp$ at the Fermilab Tevatron~\cite{Cheung:2002gd}, also via $pp\to W^{\pm\ast} Z^\ast X \to H^\pm X$~\cite{Asakawa:2005gv}
and the decay process $H^\pm\to W^\mp Z$~\cite{Kanemura:1997ej} at the CERN LHC, 
and further via the processes  $e^+e^-\to W^\mp
H^\pm$~\cite{Godbole:1994np,Cheung:1994rp,Kanemura:1999tg} and
$e^+e^-\to \nu \bar \nu W^{\pm\ast} Z^\ast \to \nu \bar \nu
H^\pm$~\cite{Kanemura:2000cw} at the ILC.   
Another striking feature of models with complex isospin-triplets, such
as the GM model, the left-right symmetric model, the littlest Higgs
model, and some models motivated by neutrino masses, 
is the appearance of doubly-charged states $H^{\pm\pm}$.
At hadron colliders, such doubly-charged Higgs bosons are studied via
the pair production mechanism~\cite{Akeroyd:1998zr, Gunion:1989in, Dion:1998pw} as well as the
single
production mechanism~\cite{Dion:1998pw, Akeroyd:2005gt} and the $W$-boson fusion
mechanism~\cite{Vega:1989tt, Huitu:2000ut}. 
They can also be investigated at the ILC and its $e^-e^-$,
$e^-\gamma$ and $\gamma\gamma$ option in various scenarios~\cite{Gunion:1998ii}.

%
\section{The S-matrix for two-body elastic scatterings}
%

In this section, we calculate the transition matrix of elastic
scatterings of two scalar-boson states in the GM model.
The transition matrix $T(\varphi_1 \varphi_2 \to \varphi_3 \varphi_4)$
is equivalent to the S-wave 
amplitude $\langle \varphi_3 \varphi_4 | a^0 | \varphi_1 \varphi_2
\rangle$
at high energies ($\sqrt{s}\gg m_W^{2}$), where $\varphi_i$ represent
longitudinally-polarized weak 
bosons or physical Higgs bosons of the model.
The condition of partial wave unitarity is given
for the S-wave amplitude matrix 
by~\cite{Gunion:1992hs,Lee:1977yc}
\bea
| \langle \varphi_3 \varphi_4 | a^0 |\varphi_1 \varphi_2 \rangle | < \frac{1}{2}.  
\label{PU}
\eea
We employ this condition in the high energy limit to constrain 
the model parameters in the next section.
Thanks to the equivalence theorem~\cite{Cornwall:1974km}, the S-matrix elements in which
longitudinally-polarized weak bosons are in initial and final states
are equivalent to those in which these weak bosons are replaced by the 
corresponding Nambu-Goldstone bosons in the high energy limit~\cite{ Lee:1977yc}.
In addition, in this limit, only quartic couplings (scalar contact
interactions) of the
Higgs-Goldstone couplings are relevant to the unitarity conditions, which
can be translated into the bounds on the related Higgs-boson masses after Eq.(\ref{PU}) is imposed.
Therefore, we here  evaluate the matrix  $\langle \varphi_3 \varphi_4 | a^0 |\varphi_1 \varphi_2 \rangle$
taking into account all possible two-body scalar channels in the high energy limit, and obtain all the eigenstates and the eigenvalues.

Under $O(4) (\simeq SU(2)_L \otimes SU(2)_R)$, the field components of
$\phi$, $\chi$ and $\xi$ are
expressed by a \underline{4} and a \underline{9} representations as  
\bea
\Psi_D &=&\left(
\omega_1, \omega_2, \phi_r^0,  \phi_i^0 
\right), \\
\Psi_T &=&\left(
\chi_1, \chi_2,  \chi_3,  \chi_4,  \xi_1,  \xi_2, \chi_r^0,  \chi_i^0,  \xi_r^0 
\right),
\eea
where $\phi^+ = (\omega_1 + i \omega_2)/\sqrt{2}$, $\phi^{0}=(\phi_r^0 + i \phi_i^0)/\sqrt{2}$, 
$\chi^{++}=(\chi_1 +
i \chi_2)/\sqrt{2}$, $\chi^{+}=(\chi_3 + i \chi_4)/\sqrt{2}$,
$\chi^{0}=(\chi_r^0 + i \chi_i^0)/\sqrt{2}$ 
and $\xi^{+}=(\xi_1 + i \xi_2)/\sqrt{2}$. 
We consider all possible two-body scattering channels ($\Psi_{a} \Psi_{b}
\to \Psi_{c} \Psi_{d}$) not only for the
neutral two-body states as initial and final states but also for  
the singly-, the doubly-, the triply- and the
quadruply-charged two-body states. 
There are totally 91 initial (or final) two-body states, in which 25 are the neutral, 36 are singly-charged, 
22 are doubly-charged, 6 are triply-charged, and the last 2 are the quadruply-charged states. 
We construct the 91 $\times$ 91 transition matrix of high-energy S-wave
amplitudes, and then calculate their eigenvalues.


The initial (final) two-body states can be treated separately as $\Psi_D
\Psi_D$, $\Psi_T \Psi_T$ and $\Psi_D \Psi_T$.
The high-energy S-wave amplitudes are block-diagonalized by the electric charge and
also the discrete $Z_2$ symmetry ($\Phi \to \Phi$ and $\Delta\to -\Delta$). 
Each submatrix with respect to the $\Psi_D\Psi_D$ or $\Psi_T\Psi_T$
states can also be classified by irreducible decomposition of
direct products of the representations for $O(4)$ as  
\bea
\underline{4}\otimes \underline{4}&=&(1)_D^{} \oplus (9) \oplus (6), \\
\underline{9}\otimes \underline{9}&=&(1)_T^{} \oplus (44)\oplus (36).
\eea
The only singlet and symmetric representations,
\bea
  (1)_{a}&=& \sum_{k=1} (\Psi^k_a)^2,\\
  (s)^{ij} &=& \Psi^i_{a} \Psi^j_a
  - \frac{1}{N_a}\sum_{k=1} (\Psi^k_a)^2,
\eea
contribute to the scatterings of our interests,
where $a=D\; {\rm or}\;
  T\, ;\,  i,j=1{\rm -}4\; {\rm and}\; s=9 \;{\rm for}\; a=D\;, {\rm or} \;
  i,j=1{\rm -}9\; {\rm and} \; s=44\;{\rm
  for}\; a=T$, and $N_D^{}=4$ and $N_T^{}=9$.
For the $\Psi_D \Psi_T$ states which
should be of the $\underline{4} \otimes \underline{9}$ representation, there
is no singlet representation so that $O(4)$ cannot help for the classification.
Furthermore, several additional discrete transformations can be used to further 
classify the states, which will be defined below.\\

\noindent
{\underline {\it Neutral channels }} \\

We outline further classification of the decomposed irreducible states for
the case of the neutral 25 two-body channels (4 for $\Psi_D\Psi_D$, 11
for $\Psi_T\Psi_T$ and 10 for $\Psi_D\Psi_T$).
For $\Psi_D\Psi_D$ states, we have 
the singlet state $(1)_D^{}$ and the three neutral elements of $(9)^{ij}$
($(9)^{33}$, $(9)^{44}$ and  $(9)^{34}$), in which the $C$ parity separates 
$(9)^{34}$ from the other states. After taking
appropriate linear combination, we obtain two separate
states under the transformation of $\phi_r \to \phi_i$ and $\phi_i \to
-\phi_r$ as $((9)^{33}\pm (9)^{44})/\sqrt{2}$.
Thus four eigenstates of the transition matrix for the neutral
$\Psi_D\Psi_D \to \Psi_D\Psi_D$ channels are obtained~\cite{Lee:1977yc}.

Next, we consider $\Psi_T\Psi_T\to\Psi_T\Psi_T$ scatterings in which
both the initial and final states are electrically neutral.
In addition to the singlet state $(1)_T$, we have 10 neutral states from
$(44)^{ij}$, in which \{ 
$(44)^{11}+(44)^{22}$,
$(44)^{33}+(44)^{44}$,
$(44)^{55}+(44)^{66}$,
$(44)^{77}$,
$(44)^{88}$ \} are the diagonal element states ($i=j$),
and \{$(44)^{35}+(44)^{46}$,
$(44)^{36}-(44)^{45}$,
$(44)^{78}$,
$(44)^{79}$,
$(44)^{89}$\} are the off-diagonal element states ($i\neq j$).
Among the diagonal element states, the linear combination
$(44)^{77}-(44)^{88}$ has different property under
the transformation of $\chi_r \to \chi_i$ and $\chi_i \to - \chi_r$.
Then, linear combinations \{
$(44)^{11}+(44)^{22}+(44)^{77}+(44)^{88}$,  
$(44)^{33}+(44)^{44}+(44)^{55}+(44)^{66}$ \} and
\{
$(44)^{11}+(44)^{22}-(44)^{77}-(44)^{88}$, 
$(44)^{33}+(44)^{44}-(44)^{55}-(44)^{66}$\} 
show different property under the transformation of
$\chi^{++}\chi^{--} \leftrightarrow \chi^0\chi^0$, 
and $\chi^{+}\chi^{-} \leftrightarrow \xi^+\xi^-$. 
The first two states have completely the same property
as that of the singlet state $(1)_T$, so that the appropriate
linear combination of these three states give the three eigenstates.
For the off-diagonal element states, we can separate them by using the $C$ parity
and the transformation of $\xi \to -\xi$, so that these states are block-diagonalized to two $2 \times 2$ submatrices and
one singlet. By diagonalizing remained $2 \times 2$ matrices, we obtain
all the eigenstates for the $\Psi_T\Psi_T \to \Psi_T\Psi_T$ channels.

In order to diagonalize all the $\Psi_D\Psi_D$ and  $\Psi_T\Psi_T$ states, 
we take linear combinations of the eigenstates
of $\Psi_D\Psi_D$ and $\Psi_T\Psi_T$ that have similar transformation
properties. Consequently, all the $\Psi_D\Psi_D$ and $\Psi_T\Psi_T$ states
are completely separated, and the eigenvalues of the transition matrix
for these channels are obtained. 

Finally, classifying the ten $\Psi_D\Psi_T$ states by using the $C$
transformation as well as some discrete transformations in a similar way
to above, we completely diagonalized $25 \times 25$ scattering matrix
for the electrically neutral two-body states.

The neutral states, $|A^0_i\rangle (i=1\sim 25)$, which give
(at most $2\times 2$) block-diagonal transition matrices are found as below:
\bea
|A^0_{1}\rangle&=&(2\phi^+\phi^-+\phi_i^0\phi_i^0+\phi_r^0\phi_r^0) /2\sqrt{2}\ , \\
|A^0_2\rangle&=&(2\chi^{++}\chi^{--}+2\chi^{+}\chi^{-}+2\xi^{+}\xi^{-}+\chi_i^0\chi_i^0+\chi_r^0\chi_r^0+\xi_r^0\xi_r^0)/3\sqrt{2}\ , \\
|A^0_3\rangle&=&(2\phi^+\phi^--\phi_i^0\phi_i^0-\phi_r^0\phi_r^0) /2\sqrt{2}\ , \\
|A^0_4\rangle&=&(-  2\chi^{++}\chi^{--}+\chi_i^0\chi_i^0+ \chi_r^0\chi_r^0)/2\sqrt{2}\ , \\
|A^0_5\rangle&=&(\phi_i^0\phi_i^0-\phi_r^0\phi_r^0)/2\ ,  \\
|A^0_6\rangle&=&(\chi^+\xi^-+ \chi^-\xi^++\sqrt{2}\chi_r^0\xi_r^0)/2\ , \\
|A^0_7\rangle&=&\phi_r^0\phi_i^0\ , \\
|A^0_8\rangle&=&(\chi^+\xi^-- \chi^-\xi^+-\sqrt{2}i\xi_r^0\chi_i^0)/(2i)\ , \\
|A^0_{9}\rangle&=&(\chi_i^0\chi_i^0- \chi_r^0\chi_r^0)/2\ , \\
|A^0_{10}\rangle&=&(2\chi^{++}\chi^{--}-\chi^+\chi^--\xi^+\xi^-+\chi_i^0\chi_i^0+ \chi_r^0\chi_r^0-2\xi_r^{0}\xi_r^{0})/3\sqrt{2}\ , \\
|A^0_{11}\rangle&=&(2\chi^{++}\chi^{--}
-4\chi^{+}\chi^{-}-4\xi^{+}\xi^{-}+\chi_r^0\chi_r^0+\chi_i^0\chi_i^0+4\xi_r^0\xi_r^0)/6\sqrt{2}
\ , \\
|A^0_{12}\rangle&=&(\chi_r^0\chi_r^0-\chi_r^0\chi_i^0)/\sqrt{2}\ , \\
|A^0_{13}\rangle&=&(\chi^+\xi^-+ \chi^-\xi^+-\sqrt{2}\chi_r^0\xi_r^0)/2 \ , \\
|A^0_{14}\rangle&=&(\chi^+\xi^-- \chi^-\xi^++\sqrt{2} i\xi_r^0\chi_i^0)/(2i)\ , \\
|A^0_{15}\rangle&=& \chi_r^0\chi_i^0\ ,  \\
|A^0_{16}\rangle&=&\left\{
2(\phi^+\chi^- +\phi^-\chi^++ \phi^+\xi^-+\phi^-\xi^+)-\sqrt{2}(\phi_r^0\chi_r^0+\phi_i^0\chi_i^0)
-2 \phi_r^0\xi_r^0\right\}/3\ , \\
|A^0_{17}\rangle&=&( \phi_r^0\chi_r^0-\phi_i^0\chi_i^0)/\sqrt{2}\ , \\
|A^0_{18}\rangle&=&( \phi_i^0\chi_r^0+\phi_r^0\chi_i^0)/\sqrt{2}\ , \\
|A^0_{19}\rangle&=&\left\{2(\phi^+\chi^- - \phi^-\chi^+- \phi^+\xi^-+\phi^-\xi^+)
+\sqrt{2}i(\phi_i^0\chi_r^0-\phi_r^0\chi_i^0)-4i\phi_i^0\xi_r^0\right\}/(6i)\ , \\
|A^0_{20}\rangle&=&\left\{ \phi^+\chi^- + \phi^-\chi^++\phi^+\xi^-+\phi^-\xi^++\sqrt{2}(\phi_r^0\chi_r^0+\phi_i^0\chi_i^0)+\phi_r^0\xi^0\right\}/3\ , \\
|A^0_{21}\rangle&=&\left\{\phi^+\chi^- - \phi^-\chi^+- \phi^+\xi^-+\phi^-\xi^++\sqrt{2}i(\phi_r^0\chi_i^0-\phi_i^0\chi_r^0)+i\phi_i^0\xi_r^0\right\}/(3i)\ , \\
|A^0_{22}\rangle&=&\left\{\phi^+\chi^- - \phi^-\chi^+-\phi^+\xi^-+\phi^-\xi^++2\sqrt{2}i( \phi_i^0\chi_r^0-\phi_r^0\chi_i^0)+4i\phi_i^0\xi_r^0\right\}/(6i)\ , \\
|A^0_{23}\rangle&=&\left\{\phi^+\chi^- + \phi^-\chi^++\phi^+\xi^-+\phi^-\xi^+-2\sqrt{2}( \phi_r^0\chi_r^0+\phi_i^0\chi_i^0)+4 \phi_r^0\xi_r^0)\right\}/6\ , \\
|A^0_{24}\rangle&=&(\phi^+\chi^- + \phi^-\chi^+- \phi^+\xi^--\phi^-\xi^+)/2\ , \\
|A^0_{25}\rangle&=&(\phi^+\chi^- - \phi^-\chi^++\phi^+\xi^--\phi^-\xi^+)/(2i)\ .
\eea
The state $(1)_D$ and the three neutral linear-combined states from $(9)^{ij}$ respectively correspond to
$|A^0_{1}\rangle$ and \{$|A^0_{3}\rangle$, $|A^0_{5}\rangle$ and
$|A^0_{7}\rangle$\}.
The state $(1)_T$ and the linear combined states from $(44)^{ij}$ correspond
to  $|A^0_{2}\rangle$ and $|A^0_{i}\rangle$ $(i=4,6,8,11-15)$, respectively.
The first eight states block-diagonalize the transition matrix to four
$2 \times 2$ submatrices, and the other seventeen states give eigenstates.
\\

\noindent
{\underline {\it Singly-charged channels}}\\

There are eighteen singly-charged states with the electric charge $+1$ among
all the two-body states. The charge conservation ensures that these
states are composed of a subset among all the states with various electric charges.
The corresponding high-energy transition matrix is consequently (block-) diagonalized by
the following states: 
\bea
|A_1^+\rangle&=&(\phi^+\phi_r^0+\phi^+\phi_i^0)/\sqrt{2}\ , \\
|A_2^+\rangle&=&(\sqrt{2}\chi^{++}\chi^-+ \chi^+\chi_r^0+\chi^+\chi_i^0)/2 \ ,\\
|A_3^+\rangle&=&(\phi^+\phi_r^0-\phi^+\phi_i^0)/\sqrt{2}\ , \\
|A_4^+\rangle&=&(\sqrt{2}\chi^{++}\xi^-+ \xi^+\chi_r^0-\xi^+\chi_i^0)/2 \ ,\\
|A_5^+\rangle&=&(\sqrt{2}\chi^{++}\chi^- -\chi^+\chi_r^0-\chi^+\chi_i^0+2\xi^+ \xi_r^0)/2\sqrt{2}\ , \\
|A_6^+\rangle&=&(\sqrt{2}\chi^{++}\xi^- - \xi^+\chi_r^0+\xi^+\chi_i^0+2\chi^+ \xi_r^0)/2\sqrt{2}\ , \\
|A_7^+\rangle&=&(\sqrt{2}\chi^{++}\chi^--\chi^+\chi_r^0-\chi^+\chi_i^0-2\xi^+\xi_r^0)/2\sqrt{2}\ , \\
|A_8^+\rangle&=&(\sqrt{2}\chi^{++}\xi^- -\xi^+\chi_r^0+\xi^+\chi_i^0-2\chi^+\xi_r^0) /2\sqrt{2}\ , \\
|A_{9}^+\rangle&=&(\xi^+\chi_r^0+\xi^+\chi_i^0)/\sqrt{2}\ , \\
|A_{10}^+\rangle&=&(\chi^+\chi_r^0-\chi^+\chi_i^0)/\sqrt{2}\ , \\
|A_{11}^+\rangle &=& (\chi^{++}\phi^--2\phi^+\xi_r^0-\chi^+\phi_r^0-\chi^+\phi_i^0-\xi^+\phi_r^0+\xi^+\phi_i^0)/3\ , \\
|A_{12}^+\rangle &=& (\phi^+\chi_r^0+\phi^+\chi_i^0+\sqrt{2}\xi^+\phi_r^0+\sqrt{2}\xi^+\phi_i^0)/\sqrt{6}\ , \\
|A_{13}^+\rangle &=& (\phi^+\chi_r^0-\phi^+\chi_i^0+\sqrt{2} \chi^+\phi_r^0 -\sqrt{2}\chi^+\phi_i^0)/\sqrt{6}\ , \\
|A_{14}^+\rangle &=& (4\chi^{++}\phi^-+4\phi^+\chi_r^0-\chi^+\phi_r^0
-\chi^+\phi_i^0-\xi^+\phi_r^0+\xi^+\phi_i^0)/6 \ ,\\
|A_{15}^+\rangle &=& (\chi^+\phi_r^0+\chi^+\phi_i^0-\xi^+\phi_r^0+\xi^+\phi_i^0)/2\ , \\
|A_{16}^+\rangle &=& (\sqrt{2}\phi^+\chi_r^0+\sqrt{2}\phi^+\chi_i^0- \xi^+\phi_r^0-\xi^+\phi_i^0)/\sqrt{6}\ , \\
|A_{17}^+\rangle &=& (\sqrt{2}\phi^+\chi_r^0-\sqrt{2}\phi^+\chi_i^0- \chi^+\phi_r^0 +\chi^+\phi_i^0)/\sqrt{6}\ , \\
|A_{18}^+\rangle &=& (2\chi^{++}\phi^--\phi^+\xi_r^0+\chi^+\phi_r^0
+\chi^+\phi_i^0+\xi^+\phi_r^0-\xi^+\phi_i^0)/3\ . 
\eea
The eighteen singly-charged states with the electric charge $-1$ can be obtained
by the $C$ transformation for the above states with the charge $+1$.\\

\noindent
{\underline {\it Doubly-charged channels}}\\

There are eleven doubly-charged two-body states with the electric charge
$+2$. We can decompose the subset of the transition matrix for these
states to at most $2 \times 2$ matrices by the following linear combination;    
\bea
|A_1^{++}\rangle&=&\phi^+\phi^+, \\
|A_2^{++}\rangle&=&(\chi^+\xi^+ -\chi^{++}\xi_r^0)/\sqrt{2}\ , \\ 
|A_3^{++}\rangle&=&\chi^+\chi^+, \\
|A_4^{++}\rangle&=&(\chi^{++}\chi_r^0 -\chi^{++}\chi_i^0)/\sqrt{2}\ , \\ 
|A_5^{++}\rangle&=&\xi^+\xi^+, \\
|A_6^{++}\rangle&=&(\chi^{++}\chi_r^0 +\chi^{++}\chi_i^0)/\sqrt{2}\ , \\ 
|A_7^{++}\rangle&=&(\chi^+\xi^+ +\chi^{++}\xi_r^0)/\sqrt{2}\ , \\ 
|A_8^{++}\rangle&=&(\phi^+\chi^+ +\phi^+\xi^+   +\chi^{++}\phi_r^0)/\sqrt{3}\ , \\ 
|A_9^{++}\rangle&=&(\phi^+\chi^+ -\phi^+\xi^+   -\chi^{++}\phi_i^0)/\sqrt{3}\ , \\ 
|A_{10}^{++}\rangle&=&(\phi^+\chi^+ +\phi^+\xi^+   -2\chi^{++}\phi_r^0)/\sqrt{6}\ , \\
|A_{11}^{++}\rangle&=&(\phi^+\chi^+ - \phi^+\xi^+  +2\chi^{++}\phi_i^0)/\sqrt{6}\ . 
\eea
The corresponding doubly-charged two-body states with the charge $-2$ can be
obtained by $C$ transformation in the above states with the charge $+2$. \\

\noindent
{\underline {\it Triply-charged channels}}\\

There are three triply-charged states with the electric charge $+3$, and
the subset of the transition matrix for the initial and final states
can be diagonalized by the following eigenstates as 
\bea
|A_1^{+++}\rangle&=&\chi^{++}\phi^+, \\
|A_2^{+++}\rangle&=&\chi^{++}\chi^+, \\
|A_3^{+++}\rangle&=&\chi^{++}\xi^+. 
\eea
All the eigenstates with the opposite electric charge can be obtained
by the $C$ transformation of these eigenstates with the charge $+3$.\\

\noindent
{\underline {\it Quadruply-charged channels}}\\

Finally, we have only one quadruply-charged state for each electric
charge of $+4$ and $-4$,  
\bea
|A_1^{++++}\rangle&=&\chi^{++}\chi^{++}, \\
|A_1^{----}\rangle&=&\chi^{--}\chi^{--}. 
\eea

\noindent
{\underline {\it Eigenvalues of the transition matrix for all channels}}\\

In summary, the transition matrix $T$ has been block-diagonalized
as
\bea
 T = \left[
      \begin{array}{cccccccccc}
        T^{0}  & 0 & 0 & 0 & 0 & 0 & 0 & 0 & 0 \\
        0 & T^{+}  & 0 & 0 & 0 & 0 & 0 & 0 & 0 \\
        0 & 0 & T^{-}  & 0 & 0 & 0 & 0 & 0 & 0\\
        0 & 0 & 0 & T^{++} & 0 & 0 & 0 & 0 & 0\\
        0 & 0 & 0 & 0 &T^{--}  & 0 & 0 & 0 & 0\\
        0 & 0 & 0 & 0 & 0 & T^{+++}& 0 & 0 & 0\\
        0 & 0 & 0 & 0 & 0 & 0 & T^{---} & 0 &0  \\
        0 & 0 & 0 & 0 & 0 & 0 & 0 & T^{++++}&0  \\
        0 & 0 & 0 & 0 & 0 & 0 & 0 & 0 & T^{----}  \\
       \end{array}
     \right],
\label{Tmatrix}
\eea
where block-diagonal transition submatrices for the neutral, the singly-charged, the doubly-charged, the triply-charged,
and the quadruply-charged two-body states, $T^0$, $T^\pm$,  $T^{\pm\pm}$, $T^{\pm\pm\pm}$, and $T^{\pm\pm\pm\pm}$, respectively, are given by
\bea
T^0&=&{\rm diag}({\bf X_1}, {\bf X_2}, {\bf X_2}, {\bf X_2}, y_1, y_1, y_2, y_2, y_2, y_2, y_2, y_3, y_3, y_3, y_3,y_3, y_4, y_4, y_4, y_5, y_5) \ , ~~~\\
T^\pm&=&{\rm diag}({\bf X_3}, {\bf X_3}, y_6,y_6, y_7,y_7, 
y_2, y_2,
y_3, y_3, y_3, y_4, y_4, y_4, y_4, y_5) \ , \\
T^{\pm\pm}&=&{\rm diag}({\bf X_4}, {\bf X_5}, {\bf X_5}, y_2, y_3, y_3, y_4, y_4) \ , \\
T^{\pm\pm\pm}&=&{\rm diag}( y_3, y_2,y_2) \ ,  \\
T^{\pm\pm\pm\pm}&=&2 y_2 \ .
\eea
Here ${\bf X_i}$ are the $2 \times 2$ matrices whose eigenvalues $x_i^\pm$ are given by 
\bea
x_1^\pm &=& 12\lambda_1+22\lambda_2+14\lambda_4\pm \sqrt{(12\lambda_1-22\lambda_2-14\lambda_4)^2+144\lambda_3^2}\ , 
\label{x1}\\
x_2^\pm &=& 4\lambda_1+4\lambda_2-2\lambda_4\pm \sqrt{(4\lambda_1-4\lambda_2+2\lambda_4)^2+4\lambda_5^2}\ , \\
x_3^\pm &=& 4\lambda_2+4\lambda_1\pm \sqrt{(4\lambda_2-4\lambda_1)^2+4\lambda_5^2}\ ,  \\
x_4^\pm &=& 8\lambda_1+4\lambda_2-2\lambda_4\pm \sqrt{(8\lambda_1-4\lambda_2+2\lambda_4)^2+8\lambda_5^2}\ ,  \\
x_5^\pm &=& 12\lambda_2+14\lambda_4\pm 2\sqrt{4\lambda_2^2+4\lambda_2\lambda_4+17\lambda_4^2}\ . 
\eea
The eigenvalues $y_i$ are obtained as
\bea
y_1&=&8\lambda_2+16\lambda_4\ , \\
y_2&=&8\lambda_2+4\lambda_4\ , \\
\label{b2}
y_3&=&4\lambda_3+\lambda_5\ , \\
y_4&=&4\lambda_3-2\lambda_5\  , \\
y_5&=&4(\lambda_3+\lambda_5)\ , \\
y_6&=& 8\lambda_2+4( 2+ \sqrt{2})\lambda_4 \ ,\\
y_7&=& 8\lambda_2+4( 2- \sqrt{2})\lambda_4 \ .
\label{y5}
\eea
Although the transition matrix between initial and final two-body states
is originally $91 \times 91$, the number of independent eigenvalues
turns out to be only seventeen.

\section{Unitarity bounds on the masses}

In this section we analyze mass bounds on the Higgs bosons in the GM model,
imposing the condition of perturbative unitarity in Eq. (\ref{PU})
to the transition matrix given in the previous section. Consequently we obtain seventeen inequations with
respect to all the independent eigenvalues of the transition matrix $T$ in Eq.(\ref{Tmatrix}) as 
\bea
|x_1^\pm|, |x_2^\pm|, |x_3^\pm|, |x_4^\pm|, |x_5^\pm|, 
|y_1|, |y_2|, |y_3|, |y_4|, |y_5|, |y_6|, |y_7| < 8\pi .
\label{PUconstraint}
\eea
These eigenvalues are respectively given in Eqs. (\ref{x1}) - (\ref{y5}) as
a combination of the dimensionless coupling constants $\lambda_i$ ($i=1\sim5$) 
in the Higgs potential, and $\lambda_i$ are related to the Higgs boson masses 
through Eqs.(\ref{lambda1})-(\ref{lambda5}), these constraints can be translated into the bounds on
the masses $m_{\tilde H_1^0}$, $m_{\tilde H_1^{0'}}, m_{H_3}^{}$ and $m_{H_5}^{}$ 
and on the mixing angles $\theta_H$ and $\alpha$.

We here show the numerical results on the Higgs mass bounds. 
Fig.~1 shows the allowed regions of the masses in the $m_{H_3}^{}$ - $m_{H_5}^{}$ plane (a), in the $m_{H_3}^{}$ - $m_{\tilde H_1^0}$ plane (b) and 
in the $m_{\tilde H_1^0}$ - $m_{\tilde H_1^{0'}}$ plane (c). 
We vary the Higgs boson masses in the range $m_{H_3}^{}, m_{H_5}^{}, m_{\tilde H_1^0}, m_{\tilde H_1^{0'}}  <$ 1 TeV and the mixing angles 
for $0 < \theta_H \le\pi/2$ and $-\pi/2 < \alpha \le \pi/2$. 
In each figure, the conditions of perturbative unitarity in Eq.(\ref{PU}) are satisfied inside the regions.
In Fig.~1(a), light shadowed region is excluded by the $Z\to b\bar b$ result.

\begin{figure}[ht]
\begin{center}
\includegraphics[width=4.9cm]{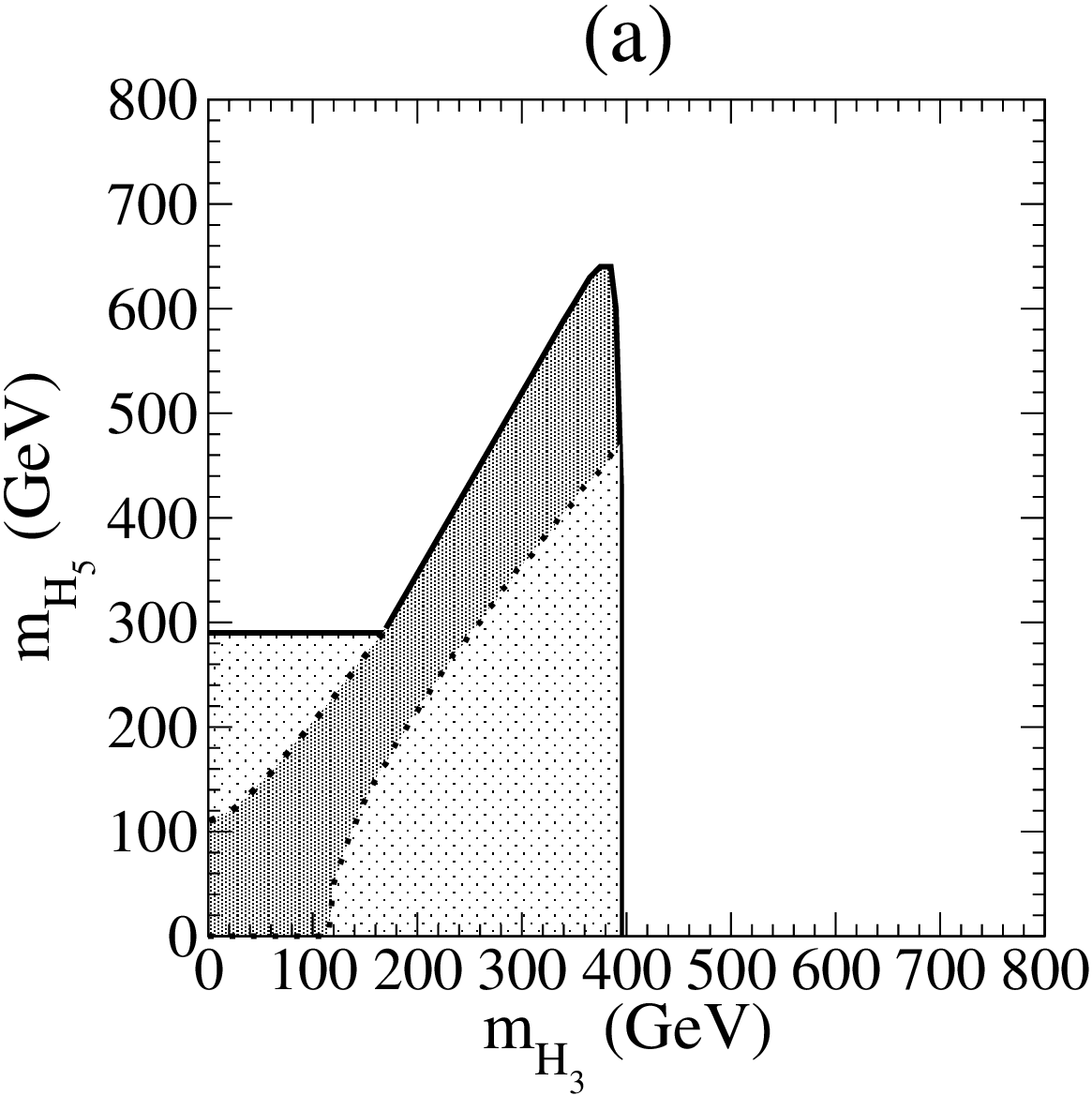}~~~~
\includegraphics[width=4.9cm]{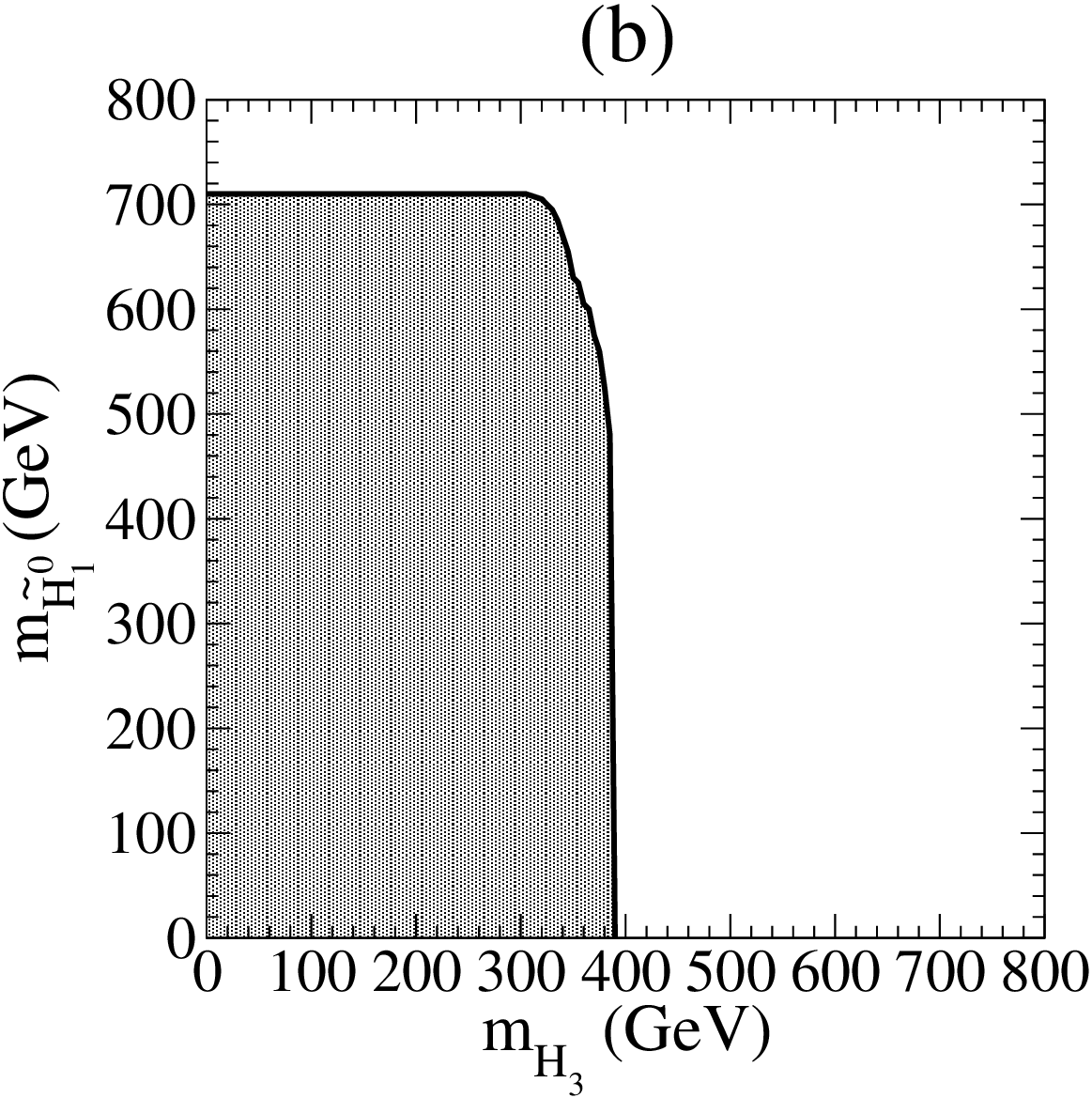}~~~~
\includegraphics[width=4.9cm]{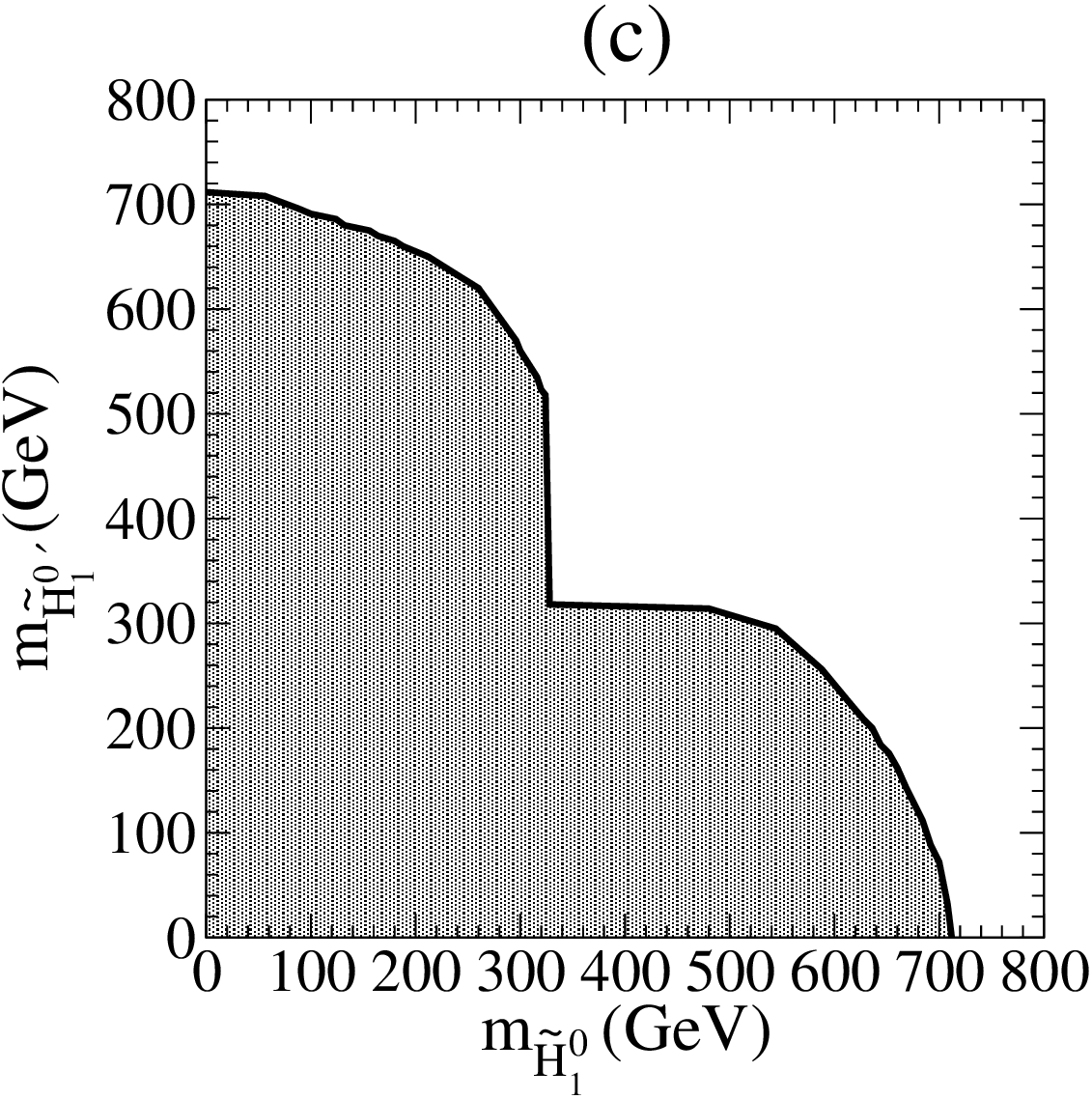}
\caption{%
Allowed regions of the masses of the Higgs bosons in the $m_{H_3}^{}$ -
 $m_{H_5}^{}$ plane (a), in the $m_{H_3}^{}$ - $m_{\tilde H_1^0}$ plane (b)
 and in the $m_{\tilde H_1^0}$ - $m_{\tilde H_1^{0'}}$ plane (c). In Fig.~1(a)
 the light shadowed regions are excluded by the $Zb\bar b$ results. 
}
\label{figure}
\end{center}
\end{figure}

Fig.~1(a) shows that $m_{H_3}^{}$ and $m_{H_5}^{}$ are bounded from above respectively by about 400 GeV and about 700 GeV. 
These upper bound come from $|x_1^\pm| < 8\pi$, which give the most stringent constraint 
among the inequations in Eq.(\ref{PUconstraint}). 
For $m_{H_3}^{} \gsim 170$ GeV, $m_{H_5}$ is bounded from above whose border is approximately corresponding 
to $\lambda_4 \simeq 0$ or $m_{H_5}^{}\simeq \sqrt{3} m_{H_3}^{}$ with $\theta_H \simeq 0$.
Owing to the factor $\sqrt{3}$, the more strict constraint is given on $m_{H_3}^{}$ than on $m_{H_5}^{}$.  
For $m_{H_3}^{} \lsim 170$ GeV, on the other hand,
the upper bound on $m_{H_5}^{}$ (290 GeV) is realized at $\theta_H\simeq \pi/2$.
When all masses other than $m_{H_3}^{}$ are zero, $m_{H_3}^{}$ is bounded by 
$m_{H_3}<\sqrt{2\pi/3}\, v$ ($\simeq 356$ GeV) from $|x_1^\pm| < 8\pi$.
However numerical analysis shows that the actual upper bound is
a few decade GeV greater. 
This excess comes from some delicate cancellation in Eq.(\ref{x1}).
When we impose the experimental data from $Z\to b\bar b$ which give the
constraint on the combination     
of $m_{H_3}^{}$ and $\theta_H$~\cite{Haber:1999zh},
the allowed region is further limited in the dark shadowed regions.
The upper bounds on $m_{H_3}^{}$ and $m_{H_5}^{}$ do not change, but the remained allowed region is
in the vicinity of $m_{H_5}^{}\simeq \sqrt{3} m_{H_3}^{}$.

In Fig.~1(b) we can see that the upper bound on $m_{\tilde H_1^0}$ is about 710 GeV, which is almost the same as that on the mass of the SM Higgs boson~\cite{Lee:1977yc}.
Larger values of $m_{\tilde H_1^0}$ are allowed for smaller values of
$\alpha$ and $\theta$ as well as smaller $m_{H_5}^{}$ and $m_{\tilde
H_1^{0'}}$ values.
For instance, taking the limit $\alpha \to 0,~ \theta_H \to 0,~m_{H_5}^{}\to 0$ and $m_{\tilde H_1^{0'}}\to 0$,
we obtain $m_{\tilde H_1^0} < \sqrt{8\pi/3}\, v$ ($\simeq 712$ GeV) 
in the condition $|x_1^+|<8\pi$.
The mass bound for another singlet $\tilde H_1^{0'}$ can be obtained by 
replacing $m_{\tilde H_1^0}$ with $m_{\tilde H_1^{0'}}$ and $\alpha$ with $\alpha + \pi/2$, which can be seen from Eqs.(\ref{lambda1})-(\ref{lambda3}).
Consequently the allowed regions in the $m_{H_3}^{}$ - $m_{\tilde H_1^{0'}}$ plane 
are given by the same as in Fig.~1(b).
We find that contrary to the result in Fig.~1(a) there is only few difference in the case where we include
the $Z\to b\bar b$ data.

The allowed region in the $m_{\tilde H_1^{0}}$ - $m_{\tilde H_1^{0'}}$
plane in Fig.~1(c) is symmetrical about the line of $m_{\tilde H_1^{0}} = m_{\tilde H_1^{0'}}(\equiv m)$. 
It is notable that at least one singlet receives very strict constraint from perturbative 
unitarity. 
The mass of lighter singlet, either $\tilde H_1^0$ or $\tilde H_1^{0'}$, is bounded from above by 322 GeV. 
In analytic calculation, this upper bound is obtained as 
$m < \sqrt{6\pi/11}\, v$ ($\simeq 322$ GeV) from $|x_1^\pm| < 8\pi$.

In the following, 
we evaluate the upper bound on 
$m_{\rm lightest}^{}$ $\left[\equiv {\rm Min}\left(m_{H_3}^{}, m_{H_5}^{}, m_{\tilde H_1^0}, m_{\tilde H_1^{0'}} \right)\right]$.
We start from the case in which the constraints from the $Zb\bar b$ results are switched off.
When the 3-plet is the lightest, we obtain the upper bound on $m_{\rm lightest} ( = m_{H_3}^{})$ as
\bea
m_{\rm lightest} < 269 {\rm~GeV},
\label{lightest}
\eea
which is considerably lower than that of the SM Higgs boson, 712 GeV.
This condition comes from the constraints $|x_1^+|<8\pi$ and $|x_5^-|<8\pi$.
Similar analysis has been done for $m_{\rm lightest}^{} = m_{H_5}^{},
m_{\tilde H_1^0}$ and $m_{\tilde H_1^{0'}}$ in order, 
and the same bound as in Eq.(\ref{lightest}) is derived for each case. 
When $m_{\rm lightest}^{} \simeq 269$ GeV, all the masses are degenerate in mass
($m_{H_3}^{} = m_{H_5}^{} = m_{\tilde H_1^0} = m_{\tilde H_1^{0'}}$).
The situation turns out to be quite similar to the situation of the two-Higgs-doublet model with 
the discrete symmetry, where the lightest of all Higgs masses are bounded at 410 GeV~\cite{Kanemura:1993hm}.
In the case of the GM model the number of the two-body states is greater
than that in the two-Higgs-doublet model. (The neutral two body
states are 14 channels in the two-Higgs-doublet model and 25 channels in
the GM model.)
Thereby we have obtained the stronger bounds than the two-Higgs-doublet model.
Finally, when we take into account the $Zb\bar b$ results~\cite{Haber:1999zh}, 
the angle $\theta_H$ is more limited for smaller values of
$m_{H_3}^{}$. Consequently, the combined upper bound on 
$m_{\rm lightest}^{}$ becomes lower than 269 GeV.
Depending on what the lightest is, the combined upper bound
turns out to be about 249 GeV (176 GeV) when
$m_{H_3}^{}$, $m_{\tilde H_1^0}$ or $m_{\tilde H_1^{0'}}$ ($m_{H_5}^{}$)
is the lightest.

We have not included the LEP direct search results, which give the lower bound 
$m_{H_{\rm SM}}^{}>114$ GeV
in the SM~\cite{Yao:2006px}. In the GM model similar lower mass bounds
can be obtained for neutral Higgs bosons but depending on the mixing
angles, which would slightly affect the upper bounds by using the
results in Figs.~1(a), (b) and (c).
We have taken into account only the $Z\to b\bar b$ result
as the experimental constraint~\cite{Haber:1999zh}, because this
constraint drastically changes the bound in the $m_{H_3}^{}$-$m_{H_5}^{}$
plane and also that on $m_{\rm lightest}$. 

\section{Conclusions}

In this paper, we have analyzed unitarity constraints on the Higgs boson
masses in the GM model, which includes a real and a complex isospin triplet
fields but predicts $\rho=1$ at tree level.
All possible two-body elastic-scattering channels (91 channels) have been taken into
account to construct the S-wave amplitude matrix in the high energy limit.
The condition of S-wave unitarity in Eq.(\ref{PUconstraint}) has been applied to the
eigenmatrix.  

We have found that all the Higgs bosons receive their masses
from the VEV under the discrete $Z_2$ symmetry, so that all the masses can be bounded from above by the condition of parturbative unitarity. 
In particular, the upper bound on the mass of the $SU(2)_{\rm V}$ 3-plet is
about $1/\sqrt{3}$ lower than that on the SM Higgs boson mass (712 GeV).
Hence at least one of the singly-charged Higgs boson masses is bounded from above at about 400 GeV.
The mass of the lighter $SU(2)_{\rm V}$ singlet scalar state, either $\tilde{H}_0$ or
$\tilde{H}_0'$, turns out to be bounded from above by about 300 GeV.  
Furthermore, the mass of the lightest Higgs boson among the 5-plet, the 3-plet and
the two singlet
states, whatever it would be, receives very strong constraint from
above; i.e., $m_{\rm lightest}^{} \simeq 270$ GeV. The point of the parameter
space at which $m_{\rm lightest}^{}$ takes its maximum value
corresponds to that where all the mass parameters are degenerate.  
The combined upper bound with the $Zb\bar b$ results becomes about 150 GeV (95\% C.L.).
Therefore, the model turns out to be well testable at collider
experiments.
The $SU(2)_{\rm V}$ 5-plet and 3-plet have the doubly- and singly-charged
states,  so that the distinctive phenomenological
features of this model should also appear in physics of charged Higgs bosons. 
Detailed phenomenological features will be discussed elsewhere.

In the analysis above, we have considered the Higgs potential with the $Z_2$
symmetry, neglecting the trilinear scalar terms of
$\mu_1$ and $\mu_2$.
The imposition of the $Z_2$ symmetry in our analysis would be justified
to avoid large excess of the neutrino masses.
When we do not respect the $Z_2$ symmetry,
the upper bounds in above results become relaxed according to the
scales of $\mu_1$ and $\mu_2$ which have linear mass dimension.
Unless $\mu_1$ and $\mu_2$ are substantially larger than ${\cal O}$(100) GeV, our 
results above can sufficiently be applied by small relaxation. 

Finally, in this paper, we have employed partial wave unitarity to constrain
parameters of the GM model at the tree level.
A more detailed study with the radiative effects such as vacuum
stability or triviality might give more strict bounds
on the Higgs boson masses in this model.

\section*{Acknowledgments}
The work of M.~A. was supported, in part, by Japan Society for the Promotion of Science. 
The work of S.~K. was supported, in part, by Grant-in-Aid of the Ministry of Education, 
Culture, Sports, Science and Technology, Government of Japan, No. 18034004, and by Grant-in-Aid for Scientific Research, Japan Society for the Promotion of Science, No. 19540277.



\begin{thebibliography}{99}

\bibitem{Gildener:1976ih}
  E.~Gildener and S.~Weinberg,
  Phys.\ Rev.\  D {\bf 13}, 3333 (1976).
  
\bibitem{Gunion:1992hs}
  J.~F.~Gunion, H.~E.~Haber, G.~L.~Kane and S.~Dawson,
  {\it The Higgs Hunter's Guide}, (Addison-Wesley, New York, 1990),
  arXiv:hep-ph/9302272.

 
 
\bibitem{Georgi:1985nv}
  H.~Georgi and M.~Machacek,
  Nucl.\ Phys.\  B {\bf 262}, 463 (1985).

\bibitem{Chanowitz:1985ug}
  M.~S.~Chanowitz and M.~Golden,
  Phys.\ Lett.\  B {\bf 165}, 105 (1985).


\bibitem{Gunion:1989ci}
  J.~F.~Gunion, R.~Vega and J.~Wudka,
  Phys.\ Rev.\  D {\bf 42}, 1673 (1990).

\bibitem{Vega:1989tt}
  R.~Vega and D.~A.~Dicus,
  Nucl.\ Phys.\  B {\bf 329}, 533 (1990).

\bibitem{Gunion:1990dt}
  J.~F.~Gunion, R.~Vega and J.~Wudka,
  Phys.\ Rev.\  D {\bf 43}, 2322 (1991).

\bibitem{Godbole:1994np}
  R.~Godbole, B.~Mukhopadhyaya and M.~Nowakowski,
  Phys.\ Lett.\  B {\bf 352}, 388 (1995)
  [arXiv:hep-ph/9411324].


\bibitem{Cheung:1994rp}
  K.~Cheung, R.~J.~N.~Phillips and A.~Pilaftsis,
  Phys.\ Rev.\  D {\bf 51}, 4731 (1995)
  [arXiv:hep-ph/9411333].
 
\bibitem{Akeroyd:1998zr}
  A.~G.~Akeroyd,
  Phys.\ Lett.\  B {\bf 442}, 335 (1998)
  [arXiv:hep-ph/9807409].


\bibitem{Haber:1999zh}
  H.~E.~Haber and H.~E.~Logan,
  Phys.\ Rev.\  D {\bf 62}, 015011 (2000)
  [arXiv:hep-ph/9909335].


\bibitem{Cheung:2002gd}
  K.~Cheung and D.~K.~Ghosh,
  JHEP {\bf 0211}, 048 (2002)
  [arXiv:hep-ph/0208254].


\bibitem{Lee:1977yc}
  B.~W.~Lee, C.~Quigg and H.~B.~Thacker,
  Phys.\ Rev.\ Lett.\  {\bf 38}, 883 (1977); 
  Phys.\ Rev.\  D {\bf 16}, 1519 (1977).

\bibitem{Dicus:1992vj}
  D.~A.~Dicus and V.~S.~Mathur,
  Phys.\ Rev.\  D {\bf 7}, 3111 (1973).


\bibitem{Lindner:1985uk}
  M.~Lindner,
  Z.\ Phys.\  C {\bf 31}, 295 (1986);
  N.~Cabibbo, L.~Maiani, G.~Parisi and R.~Petronzio,
  Nucl.\ Phys.\  B {\bf 158}, 295 (1979).

\bibitem{Kanemura:1993hm}
  S.~Kanemura, T.~Kubota and E.~Takasugi,
  Phys.\ Lett.\  B {\bf 313}, 155 (1993)
  [arXiv:hep-ph/9303263].

\bibitem{Ginzburg:2005dt}  H.~H\"{u}ffel and G.~Pocsik, 
                      Z.\ Phys.\ C {\bf 8} (1981) 13; 
                      J.~Maalampi, J.~Sirkka and I.~Vilja, 
                      \Journal{\PLB}{265}{371}{1991};  
                      A.~Akeroyd, A.~Arhrib, E.-M.~Naimi, 
                      \Journal{\PLB}{490}{119}{2000};
                      I.~F.~Ginzburg, I.~P.~Ivanov, hep-ph/0312374.
  I.~F.~Ginzburg and I.~P.~Ivanov,
  Phys.\ Rev.\  D {\bf 72}, 115010 (2005)
  [arXiv:hep-ph/0508020];
  
\bibitem{Komatsu:1981xh}
  H.~Komatsu,
  Prog.\ Theor.\ Phys.\  {\bf 67}, 1177 (1982);
R.A.~Flores and M.~Sher, Ann.\ Phys.\ (NY), 148 (1983) 295;
  M.~Sher,
  Phys.\ Rept.\  {\bf 179}, 273 (1989);
  D.~Kominis and R.~S.~Chivukula,
  Phys.\ Lett.\  B {\bf 304}, 152 (1993)
  [arXiv:hep-ph/9301222];
\bibitem{Nie:1998yn}
  S.~Nie and M.~Sher,
  Phys.\ Lett.\  B {\bf 449}, 89 (1999)
  [arXiv:hep-ph/9811234].
  S.~Kanemura, T.~Kasai and Y.~Okada,
  Phys.\ Lett.\  B {\bf 471}, 182 (1999)
  [arXiv:hep-ph/9903289].

\bibitem{Forshaw:2003kh}   
  J.~R.~Forshaw, A.~Sabio Vera and B.~E.~White,
  JHEP {\bf 0306}, 059 (2003)
  [arXiv:hep-ph/0302256].


\bibitem{Glashow:1976nt}
  S.~L.~Glashow and S.~Weinberg,
  Phys.\ Rev.\  D {\bf 15}, 1958 (1977).

\bibitem{Chun:2003ej}
  E.~J.~Chun, K.~Y.~Lee and S.~C.~Park,
  Phys.\ Lett.\  B {\bf 566}, 142 (2003)
  [arXiv:hep-ph/0304069];
  P.~Q.~Hung,
  Phys.\ Lett.\  B {\bf 649}, 275 (2007)
  [arXiv:hep-ph/0612004].


\bibitem{Aoki-Kanemura:neutrino}
  M.~Aoki and S.~Kanemura, Work in progress.
  

\bibitem{Kundu:1995qb}
  A.~Kundu and B.~Mukhopadhyaya,
  Int.\ J.\ Mod.\ Phys.\  A {\bf 11}, 5221 (1996)
  [arXiv:hep-ph/9507305].
  
\bibitem{Chakraverty:1995zw}
  D.~Chakraverty and A.~Kundu,
  Mod.\ Phys.\ Lett.\  A {\bf 11}, 675 (1996)
  [arXiv:hep-ph/9508234].

\bibitem{Grifols:1980uq}
  J.~A.~Grifols and A.~Mendez,
  Phys.\ Rev.\  D {\bf 22}, 1725 (1980).

\bibitem{Mukho:1990}
        
  B.~Mukhopadhyaya,
  Phys.\ Lett.\  B {\bf 252}, 123 (1990).

\bibitem{Asakawa:2005gv}
  E.~Asakawa and S.~Kanemura,
  Phys.\ Lett.\  B {\bf 626}, 111 (2005)
  [arXiv:hep-ph/0506310];
  E.~Asakawa, S.~Kanemura and J.~Kanzaki,
  Phys.\ Rev.\  D {\bf 75}, 075022 (2007)
  [arXiv:hep-ph/0612271].

\bibitem{Kanemura:1997ej}
  M.~C.~Peyranere, H.~E.~Haber and P.~Irulegui,
  Phys.\ Rev.\  D {\bf 44}, 191 (1991);
  A.~Mendez and A.~Pomarol,
  Nucl.\ Phys.\  B {\bf 349}, 369 (1991);
  S.~Kanemura,
  Phys.\ Rev.\  D {\bf 61}, 095001 (2000)
  [arXiv:hep-ph/9710237];
  A.~Arhrib, R.~Benbrik and M.~Chabab,
  J.\ Phys.\ G {\bf 34}, 907 (2007)
  [arXiv:hep-ph/0607182];
  A.~Arhrib, R.~Benbrik and M.~Chabab,
  Phys.\ Lett.\  B {\bf 644}, 248 (2007)
  [arXiv:hep-ph/0701126].
   
 \bibitem{Kanemura:1999tg}
  S.~Kanemura,
  Eur.\ Phys.\ J.\  C {\bf 17}, 473 (2000)
  [arXiv:hep-ph/9911541];  
  A.~Arhrib, M.~Capdequi Peyranere, W.~Hollik and G.~Moultaka,
  Nucl.\ Phys.\  B {\bf 581}, 34 (2000)
  [Erratum-ibid.\  {\bf 2004}, 400 (2004)]
  [arXiv:hep-ph/9912527];
  H.~E.~Logan and S.~Su,
  Phys.\ Rev.\  D {\bf 66}, 035001 (2002)
  [arXiv:hep-ph/0203270];
  O.~Brein,
  arXiv:hep-ph/0209124;
  O.~Brein and T.~Hahn,
  Eur.\ Phys.\ J.\  {\bf 52}, 397 (2007)
  [arXiv:hep-ph/0610079].

\bibitem{Kanemura:2000cw}
  S.~Kanemura, S.~Moretti and K.~Odagiri,
  JHEP {\bf 0102}, 011 (2001)
  [arXiv:hep-ph/0012030].
    
\bibitem{Gunion:1989in}
  J.~F.~Gunion, J.~Grifols, A.~Mendez, B.~Kayser and F.~I.~Olness,
  Phys.\ Rev.\  D {\bf 40}, 1546 (1989);
  J.~F.~Gunion, C.~Loomis and K.~T.~Pitts,
arXiv:hep-ph/9610237;
  G.~Azuelos, K.~Benslama and J.~Ferland,
  J.\ Phys.\ G {\bf 32}, 73 (2006)
  [arXiv:hep-ph/0503096];
  A.~Hektor, M.~Kadastik, M.~Muntel, M.~Raidal and L.~Rebane,
  Nucl.\ Phys.\  B {\bf 787}, 198 (2007)
  [arXiv:0705.1495 [hep-ph]];
  T.~Han, B.~Mukhopadhyaya, Z.~Si and K.~Wang,
  Phys.\ Rev.\  D {\bf 76}, 075013 (2007)
  [arXiv:0706.0441 [hep-ph]].


\bibitem{Dion:1998pw}
  B.~Dion, T.~Gregoire, D.~London, L.~Marleau and H.~Nadeau,
  Phys.\ Rev.\  D {\bf 59}, 075006 (1999)
  [arXiv:hep-ph/9810534].

\bibitem{Akeroyd:2005gt}
  A.~G.~Akeroyd and M.~Aoki,
  Phys.\ Rev.\  D {\bf 72}, 035011 (2005)
  [arXiv:hep-ph/0506176].

\bibitem{Huitu:2000ut}
  K.~Huitu, J.~Laitinen, J.~Maalampi and N.~Romanenko,
  Nucl.\ Phys.\  B {\bf 598}, 13 (2001)
  [arXiv:hep-ph/0006261];
  J.~Maalampi and N.~Romanenko,
  Phys.\ Lett.\  B {\bf 532}, 202 (2002)
  [arXiv:hep-ph/0201196].
 

\bibitem{Gunion:1998ii}
  J.~F.~Gunion,
  Int.\ J.\ Mod.\ Phys.\  A {\bf 13}, 2277 (1998)
  [arXiv:hep-ph/9803222];
  S.~Chakrabarti, D.~Choudhury, R.~M.~Godbole and B.~Mukhopadhyaya,
  Phys.\ Lett.\  B {\bf 434}, 347 (1998)
  [arXiv:hep-ph/9804297];
  J.~E.~Cieza Montalvo, N.~V.~.~Cortez, J.~Sa Borges and M.~D.~Tonasse,
  Nucl.\ Phys.\  A {\bf 790}, 554 (2007)
  [arXiv:hep-ph/0612039];
C.~X.~Yue, S.~Zhao and W.~Ma,
Nucl.\ Phys.\  B {\bf 784}, 36 (2007)
 [arXiv:0706.0232 [hep-ph]].


\bibitem{Cornwall:1974km}
  J.~M.~Cornwall, D.~N.~Levin and G.~Tiktopoulos,
  Phys.\ Rev.\  D {\bf 10}, 1145 (1974)
  [Erratum-ibid.\  D {\bf 11}, 972 (1975)].

\bibitem{Yao:2006px}
  W.~M.~Yao {\it et al.}  [Particle Data Group],
  J.\ Phys.\ G {\bf 33}, 1 (2006).
 
\end{thebibliography}
\end{document}